\documentclass[11pt]{aastex}
\usepackage{graphicx,emulateapj5,apjfonts}
\usepackage{onecolfloat}
\usepackage{epsf}

\shortauthors{Bell et al.}
\shorttitle{The dynamical evolution of early-type galaxies}

\newcommand{\combo}{{COMBO-17} }

\slugcomment{{\sc The Astrophysical Journal; in press }}

\begin{document}


\def\head{

\title{Dry Mergers in GEMS: \\
The Dynamical Evolution of Massive Early-Type Galaxies }

\author{Eric F.\ Bell$^1$, Thorsten Naab$^2$, 
Daniel H.\ McIntosh$^3$, Rachel S.\ Somerville$^4$, 
John A.\ R.\ Caldwell$^5$, Marco Barden$^1$, Christian Wolf$^6$, 
Hans-Walter Rix$^1$, Steven V.\ W.\ Beckwith$^{4,7}$,
Andrea Borch$^8$, Boris H\"aussler$^1$, Catherine Heymans$^{1,9}$, 
Knud Jahnke$^{10}$, Shardha Jogee$^{11}$, Sergey Koposov$^1$, 
Klaus Meisenheimer$^1$, Chien Y.\ Peng$^4$,
Sebastian F.\ Sanchez$^{12}$, \& Lutz Wisotzki$^9$}
\affil{$^1$ Max-Planck-Institut f\"ur Astronomie,
K\"onigstuhl 17, D-69117 Heidelberg, Germany; \texttt{bell@mpia.de}\\
$^2$ Universit\"ats-Sternwarte M\"unchen, Scheinerstr.\ 1, D-81679 M\"unchen, Germany \\
$^3$ Department of Astronomy, University of Massachusetts, 
710 North Pleasant Street, Amherst, MA 
01003, USA \\ 
$^4$ Space Telescope Science Institute, 3700 San Martin Drive, Baltimore MD, 21218, USA \\
$^5$ University of Texas, McDonald Observatory, Fort Davis, TX 79734, USA \\
$^6$ Department of Physics, Denys Wilkinson Bldg., University
of Oxford, Keble Road, Oxford, OX1 3RH, UK \\
$^7$ Johns Hopkins University, 9400 North Charles St., Baltimore, MD 21218, USA \\
$^8$ Astronomisches Rechen-Institut, M\"onchhofstr. 12-14, D69120 Heidelberg, 
Germany \\
$^9$ University of British Columbia,
Dept.\ of Physics and Astronomy,
6224 Agricultural Road,
Vancouver, V6T 1Z1, BC, Canada \\
$^{10}$ Astrophysikalisches Institut Potsdam, An der Sternwarte 16, D-14482 Potsdam, Germany \\
$^{11}$ Department of Astronomy, University of Texas at Austin, 1 University
	Station C1400, Austin, TX 78712-0259, USA \\
$^{12}$ Centro Astronomico Hispano Aleman de Calar Alto, C/Jesus Durban Remon 2-2, E-04004 Almeria, Spain \\
}

\begin{abstract}
We have used the $28'\times 28'$ {\it Hubble Space Telescope}
image mosaic from the GEMS (Galaxy Evolution from Morphology and 
SEDs) survey in conjunction with the \combo 
photometric redshift survey to constrain the incidence
of major mergers between spheroid-dominated
galaxies with little cold gas (dry mergers) since $z = 0.7$.  
A set of $N$-body merger simulations
was used to explore the morphological signatures of 
such interactions: they are recognizable 
either as $< 5$\,kpc separation close pairs or because of 
broad, low surface brightness tidal features and asymmetries.
Data with the depth and resolution of GEMS 
are sensitive to dry mergers between galaxies
with $M_V \la -20.5$ for $z \la 0.7$; dry mergers at higher redshifts
are not easily recovered in single-orbit {\it HST} imaging.
Six dry mergers (12 galaxies) 
with luminosity ratios between 1:1 and 4:1 were found
from a sample of 379 red early-type 
galaxies with $M_V < -20.5$ and $0.1 < z < 0.7$.  
The simulations suggest that the morphological signatures of 
dry merging are visible for $\sim 150$\,Myr and we use this timescale to 
convert the observed merger incidence into a rate.  On this basis
we find that present day spheroidal galaxies with $M_V < -20.5$ on average have undergone
between 0.5 and 2 major dry mergers since $z \sim 0.7$.

We have compared this result with the predictions of a Cold Dark
Matter based semi-analytic galaxy formation model.  The model
reproduces the observed declining major merger fraction of bright
galaxies and the space density of luminous early-type galaxies
reasonably well.
The predicted dry merger fraction is consistent with our observational
result.  Hence, hierarchical models
predict and observations now show that major dry mergers are an
important driver of the evolution of massive early-type galaxies in
recent epochs.

\end{abstract}

\keywords{galaxies: general ---  galaxies: interactions ---
galaxies: elliptical and lenticular --- galaxies: evolution --- 
galaxies: structure}
}

\twocolumn[\head]

\section{Introduction}

Numerical simulations have long predicted that
early-type galaxies, with spheroid-dominated stellar light
profiles, are a natural outcome of major galaxy mergers
\citep[e.g.,][]{toomre72,barnes96,naab03}.
Yet, both the timing and nature of the violent assembly of 
early-type galaxies remain frustratingly unclear.
Look-back studies have recently become large enough to 
demonstrate a steady growth in the 
total stellar mass in the red-sequence galaxy population 
since $z \sim 1$ \citep[e.g.,][]{chen03,bellc17,faber05}.
The majority of these red-sequence galaxies are morphologically
early-type out to at least $z \sim 0.7$ \citep{bellgems}; indeed, 
recent works have confirmed a growth in the total stellar
mass in morphologically early-type galaxies from $z \sim 1$
to the present day \citep{cross04,conselice05}.
This growth is dominated by a growing number of 
$\sim 0.5-2 L^*$ early-type galaxies 
\citep[e.g.,][]{bellc17,drory04,faber05}\footnote{Although
it is possible that the number of very massive galaxies,
with $L > 5 L^*$, does not change by a large amount since $z \sim 1.5$
\citep[e.g.,][]{saracco05}.}.
Interestingly, relatively few blue 
galaxies bright enough to be the star-forming progenitor of a massive
non-star-forming early-type galaxy are observed \citep{bellc17,faber05}.  
Thus, it has been suggested that mergers between non-star-forming
early-type galaxies may occur (so-called `dry mergers'), 
and build up the massive
early-type galaxy population \citep{bellc17,faber05}.
Such mergers are observed
at least in clustered environments \citep{vandokkum99,tran05}.
Yet, it is still unclear how important dry mergers are in driving the 
evolution of luminous, massive early-type galaxies, averaged
over all cosmic environments.

Dry mergers may also play an important role in 
explaining another interesting phenomenon:
the strong differences between the 
observed properties of high-luminosity (with $M_V \la -21$)
and low-luminosity early-type galaxies.
Luminous (i.e., massive) early-type galaxies
possess boxy isophotes, cores in their 
light distributions, and are supported primarily
by random stellar motions.  In contrast, faint early-type
galaxies have diskier isophotes, derive more
support from organized rotational motions, and
have power-law (or S\'ersic) light distributions
all the way into their central-most regions
\citep[e.g.,][]{bender88,kormendy96,lauer95,faber97,graham04}.
These structural differences may be the results of galaxy merging between
different types of progenitors \citep{khochfar03}.

There is growing evidence that the faint early-type
galaxy population has grown substantially at recent times
through galaxy merging and/or disk fading.
There is a deficit of faint red-sequence
galaxies in $z \sim 1$ galaxy clusters \citep{kodama04,delucia04a},
and those that are present appear to have rather young
stellar populations \citep{vanderwel03}.
Furthermore, faint early-type galaxies at the present day
can have important contributions from younger stars
\citep{kuntschner00,trager00,thomas04}.
Simulations of merging disks easily
yield remnants with properties consistent with
the typically disky, partially rotationally-supported
low-luminosity elliptical
galaxies \citep[e.g.,][]{toomre72,naab99,bendo00,naab03},
especially for unequal-mass interactions. 
Taken together with observations of a declining
merger rate, 
implying between 0.2 and 1 merger per $L^*$ galaxy \citep[e.g.,][although see
Lin et al.\ 2005]{lefevre00,patton02,conselice03}, 
it is likely that mergers between gas-rich galaxies have
played an important role in building up low-luminosity 
early-type galaxies.

In contrast, it is increasingly clear that mergers of gas-rich disks
cannot be the main formation route of more massive early-type
galaxies.  Not only are there relatively few blue
star-forming galaxies at $z \la 1$ luminous enough to fade into 
present-day massive non-star-forming galaxies, 
but the boxy isophotes and cores
in the light distributions of massive early-type 
galaxies are challenging to reproduce in 
mergers of disk galaxies \citep{naab01,barnes02,naab03}.
Such properties can, however, be naturally reproduced 
by mergers of elliptical galaxies \citep{naab05}.
This hypothesis is given some observational support by
\citet{vandokkum99} and \citet{tran05}, who found
that $\sim 15$\% of bright galaxies in the 
outskirts of the cluster MS1054$-$03 were in bound pairs
and were likely to merge on $\la 1$\,Gyr timescales; 
most of these are pairs of early-type galaxies.
Semi-analytic work by \citet{khochfar03,khochfar05} supports 
the formation of massive
early-type galaxies ending up in clusters by this route.  
 
A direct measure of the frequency of dry mergers
over the last half of cosmic time will
further constrain the importance of this process in
driving the evolution of the luminous early-type galaxy population.
Furthermore, because the majority of massive galaxies are red 
early-type galaxies
at all epochs since $z \sim 1$ \citep{bellc17}, the dry merger
rate gives important insight into the merging history of the most
massive galaxies at recent times.  Finally, gas-free mergers 
are relatively straightforward to model, and their luminosity
and color evolution is easily predicted, significantly simplifying 
the estimation of merger rates. 

The goal of this paper is to use these data to constrain
the incidence of dry mergers in the GEMS dataset.
To realize this goal, 
a number of ingredients must be in place.
Firstly, a scheme must be devised by which one can 
reliably and reproducibly identify mergers of gas-poor
systems.  Secondly, the timescale over which signatures of
gas-poor merging are identifiable must be characterized.
Finally, these criteria must be applied to the data; coupled
with the timescale, a merger frequency can be estimated.

In this paper, we present our first attempt to address this issue using 
high-resolution {\it HST} data from GEMS \citep[Galaxy
Evolution from Morphology and SEDs;][]{rix04},
photometric redshifts from COMBO-17 \citep[Classifying
Objects by Medium-Band Observations in 17 Filters;][]{wolf04},
and state-of-the-art $N$-body and semi-analytic simulations.
We briefly describe the data in \S
\ref{data}.  We outline the selection philosophy 
in \S \ref{phil}.
In order to devise criteria by which
one can select candidate dry mergers and understand
over which timescale these selection criteria are sensitive, 
we explore the characteristics of a suite of $N$-body simulation
mergers between spheroid-dominated galaxies  (\S
\ref{sims}).  We then apply these selection criteria
to the data in \S \ref{seln}.  We use these 
results to derive a dry merger fraction and incidence in 
\S \ref{res}.
In \S \ref{disc} we compare our results with limits on dry
merger rates from other observations, compare with published merger
rates, and compare the observations with the predictions of a Cold
Dark Matter (CDM) based semi-analytic model.  In the Appendix, we
explore the importance of high-speed fly-by (unbound) interactions,
and argue that the vast majority of tidally-distorted interacting
spheroids should be in the lowest-density environments capable of
hosting spheroid-dominated galaxies, and are therefore likely to
merge.  Throughout, we assume $\Omega_{\rm m} = 0.3$, $\Omega_{\rm
m}+\Omega_{\Lambda} = 1$, and $H_0 = 70 $\,km\,s$^{-1}$\,Mpc$^{-1}$.

\section{The Data} \label{data}

\combo has imaged the $\sim 30' \times 30'$ 
extended Chandra Deep Field South 
using 5 broad and 12 medium passbands complete to 
apparent $R$-band magnitude limits of $m_R \sim 23.5$.
Using these 17-passband photometric data in conjunction with 
galaxy, star, and AGN template spectra, classifications 
and redshifts are assigned for $\sim 99$\% of the
objects with sufficient flux.  The typical galaxy redshift accuracy
is $\delta z/(1+z) \sim 0.02$ \citep{wolf04},
allowing construction of rest-frame colors and absolute 
magnitudes accurate to $\sim 0.1$\,mag.
In order to probe interactions between 
gas-free, non-star-forming galaxies, 
we select only galaxies on the red sequence
with photometric redshifts $0.1 < z < 0.7$ for further
study, following \citet{bellc17} and \citet{mcintosh05}.  

We use F850LP imaging from the {\sc gems}
survey to provide $0\farcs07$ resolution data for 
our sample of red-selected galaxies.  
Using the Advanced Camera for Surveys \citep{ford03}
on the {\it HST}, a $\sim 28' \times 28'$ area of the Extended
Chandra Deep Field was surveyed to a depth allowing
galaxy detection to a limiting surface brightness of 
$\mu_{\rm F850LP,AB} \sim 24$\,mag\,arcsec$^{-2}$
\citep{rix04}; in practice early-type galaxies 
have relatively high surface brightnesses, and are 
limited only by the magnitude limit of 
\combo redshift classification.  

\section{The selection of dry mergers} \label{phil}

\begin{figure*}[t]
\begin{center}
\epsfxsize 18.0cm
\epsfbox{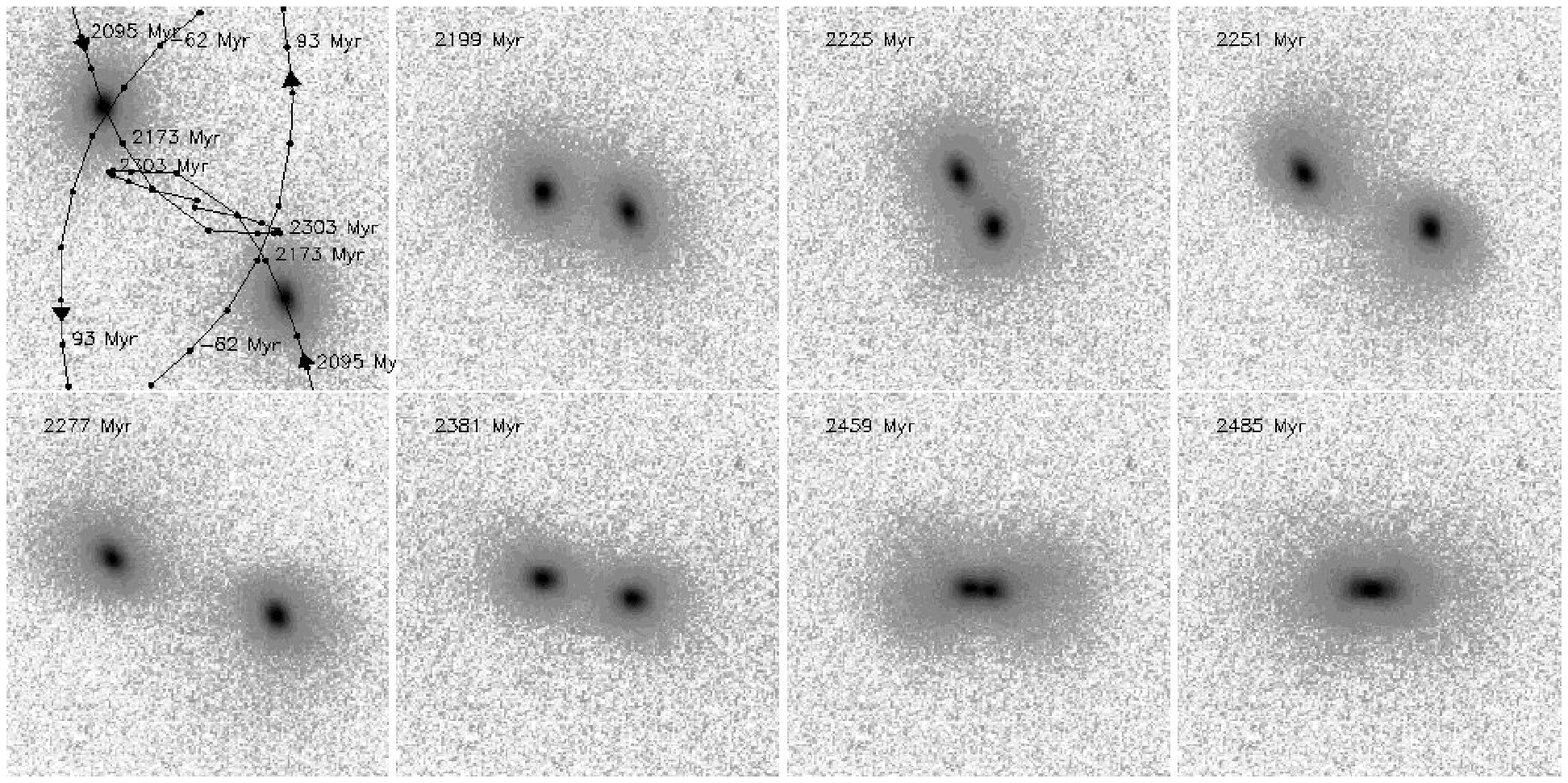}
\end{center}
\caption{\label{fig:sims} A simulation sequence showing the 
final coalescence of a gas-free elliptical--elliptical 1:1 merger.
This particular simulation had an initially parabolic orbit
with a pericenter of 21\,kpc; the face-on projection 
of the orbital plane is shown in this example.  
The merger remnant has a final $M_V = -22.5$ and 
is placed at $z = 0.5$ in real background noise from 
GEMS with the right size and apparent magnitude.  
In the first frame, the orbital paths of each galaxy 
are shown (ticks are every 26\,Myr), 
where a time of 0 corresponds to first pass, 
2.21\,Gyr to second 
pass, and final coalescence happens at 2.48\,Gyr.  
The total elapsed time between 
the first and last frame is 340\,Myr, starting 2.15\,Gyr after first pass
as the galaxies are coming towards second pass (the third frame) 
and final coalescence (the last frame of the sequence).
The images are 50\,kpc on a side.  This interaction is classifiable
as a merger for roughly 200\,Myr via 
either faint and broad tidal features or
very close double nuclei.  
}
\end{figure*}

\begin{figure*}[th]
\begin{center}
\epsfxsize 18.0cm
\epsfbox{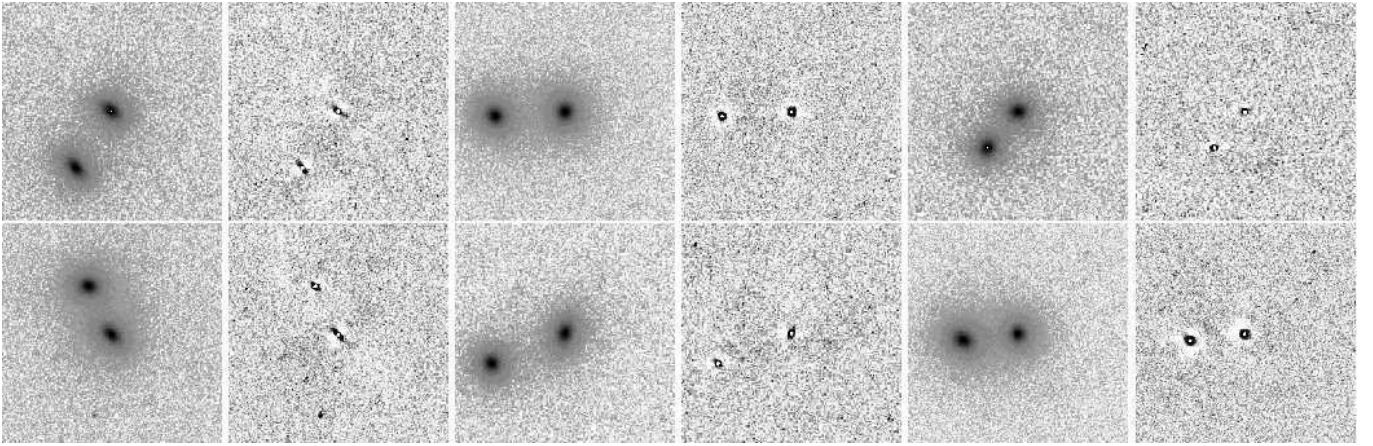}
\end{center}
\caption{\label{fig:sims_no} 
Six simulation snapshots deemed to be non-interacting 
by the majority of the classifiers.  We show the simulation 
and the GALFIT residual side-by-side for each postage stamp.
Images are 50\,kpc on a side.  The residuals in the centers
of the galaxies are from small mismatches between the 
PSF used to broaden the simulations and the GEMS PSF.
}
\end{figure*}

\begin{figure*}[th]
\begin{center}
\epsfxsize 18.0cm
\epsfbox{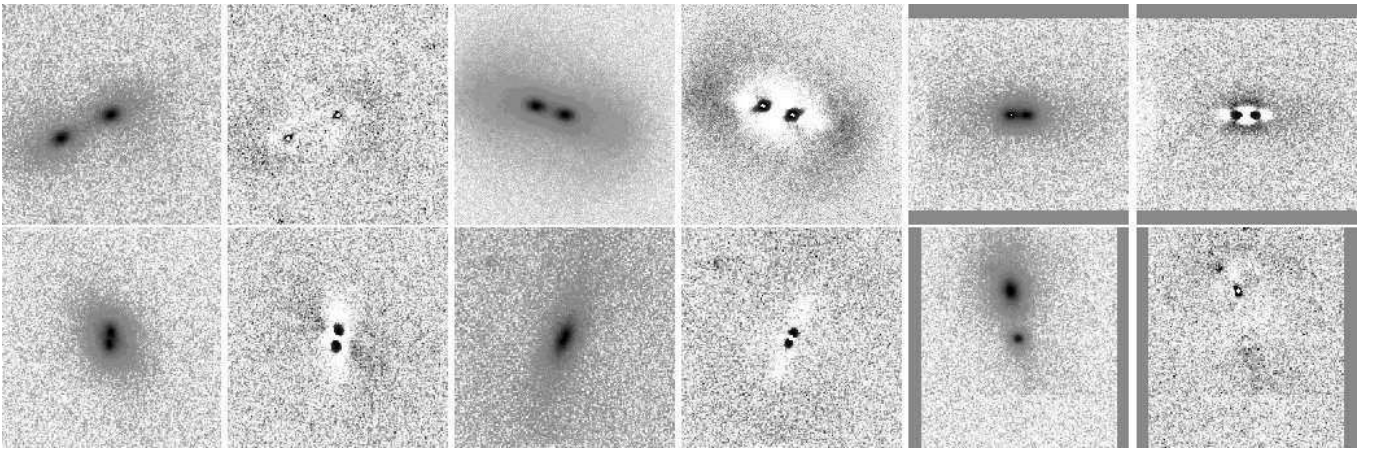}
\end{center}
\caption{\label{fig:sims_mrg} 
Six simulation snapshots deemed to be interacting
by the majority of the classifiers.  The format is similar
to Fig.\ \ref{fig:sims_no}.  In all but the top left and bottom 
right postage stamps, the merging system was determined by 
SExtractor to be one object and was therefore fit as one
object by GALFIT.  There is further discussion of this issue in 
the text.
}
\end{figure*}

\begin{figure*}[th]
\begin{center}
\epsfxsize 18.0cm
\epsfbox{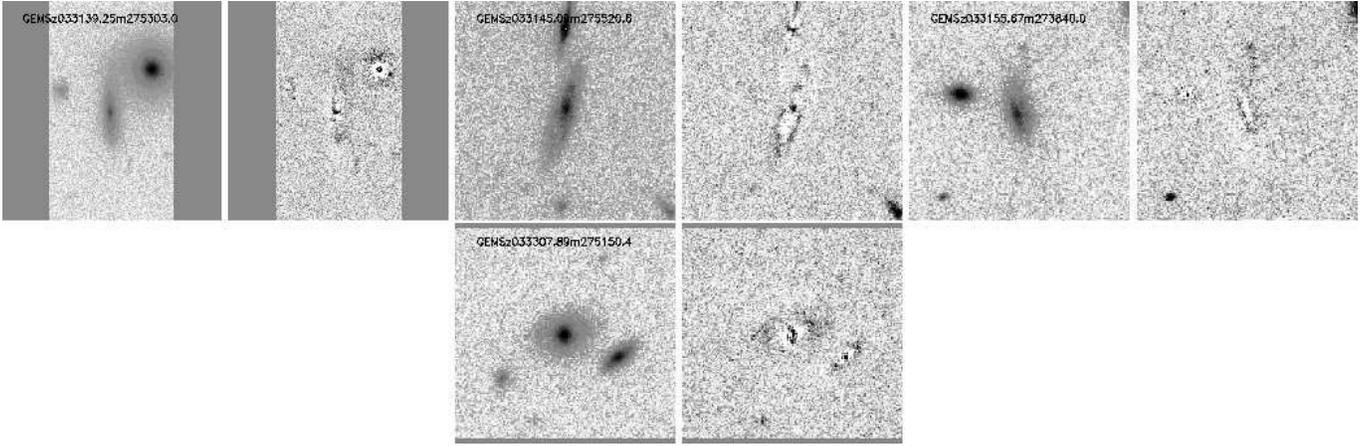}
\epsfxsize 3.0cm
\end{center}
\caption{\label{fig:nointeract} 
The four candidates which were not deemed to be interacting
by the majority of the classifiers.  We show the image
and the GALFIT residual side-by-side for each sample system.
Images are 50\,kpc on a side. 
}
\end{figure*}

\begin{figure*}[th]
\begin{center}
\epsfxsize 18.0cm
\epsfbox{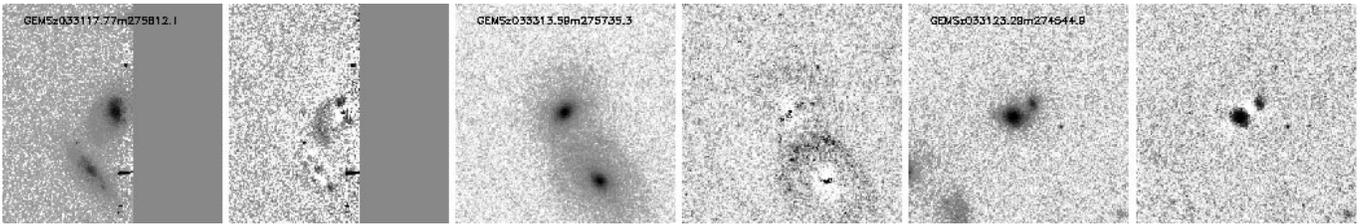}
\end{center}
\caption{\label{fig:poss} 
Three contentious candidates, arranged in order of decreasing 
star formation as judged from the classifications.
GEMS 033123.29m274544.9 was deemed by SExtractor to be one object,
and was fit by GALFIT by a single S\'ersic profile.
}
\end{figure*}

\begin{figure*}[th]
\begin{center}
\epsfxsize 18.0cm
\epsfbox{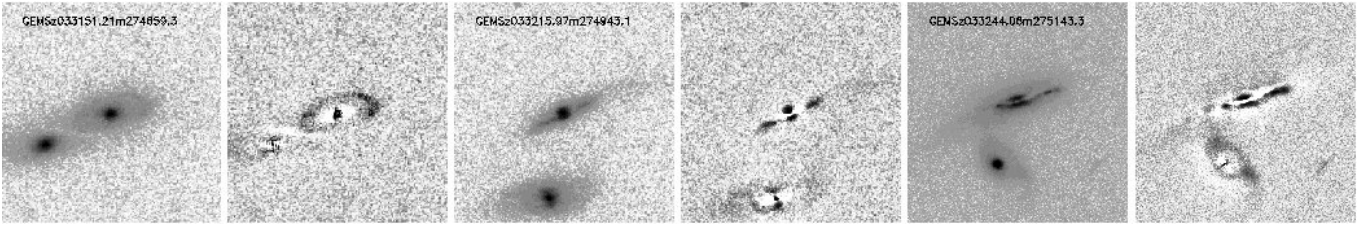}
\end{center}
\caption{\label{fig:gas_rich} Three interactions with clear signs of 
dust and/or star formation, indicating some gas content.  In 
all cases, both interaction partners have colors
consistent with the red sequence (despite oftentimes small amounts
of star formation in these systems), and substantial bulge components.
}
\end{figure*}

\begin{figure*}[th]
\begin{center}
\epsfxsize 18cm
\epsfbox{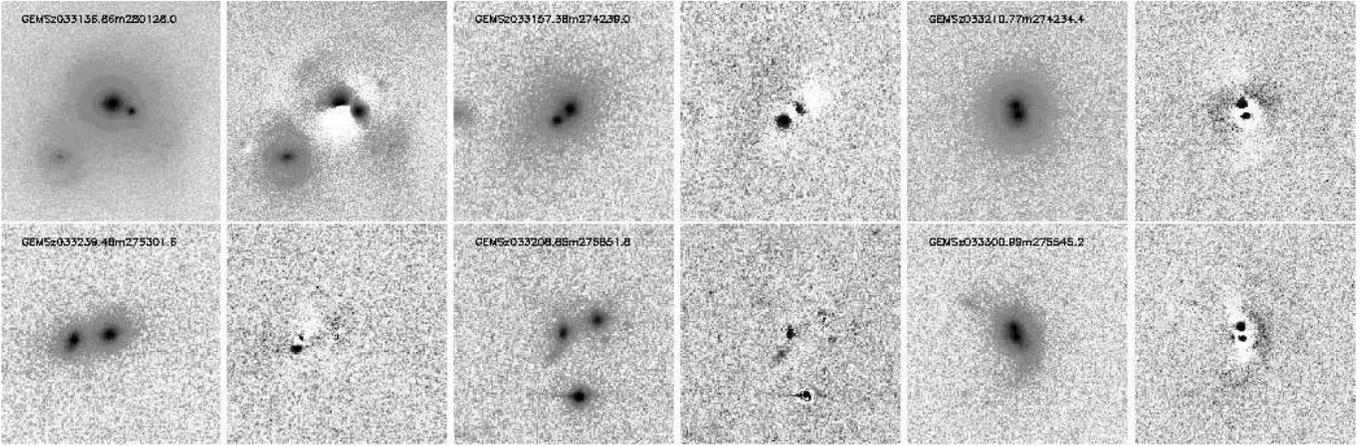}
\end{center}
\caption{\label{fig:drygals} The six candidate spheroid-dominated
merging systems in GEMS; the last 2 systems were judged by 
2 out of the 5 classifiers to have hints of ongoing star formation.  
Images are 50\,kpc on a side.  
The spiral galaxy in the lower left-hand side
of the GEMS 033136.86m280128.0 A and B postage stamp is 
a background projection.  GEMS 033136.86m280128.0, GEMS 033157.38m274239.0
GEMS 033210.77m274234.4, and GEMS 033300.99m275545.2 were 
deemed by SExtractor to be single objects,
and were therefore fit by GALFIT using single S\'ersic profiles.
}
\end{figure*}

\begin{table*}
\caption{Properties of the dry merger candidates {\label{tab}}}
\begin{center}
\begin{tabular}{lcccccc}
\hline
\hline
Galaxy name & RA & Dec & z & $M_V$ & $U - V$ & Classification\\
 & hh mm ss & dd mm ss &  &  & \\
\hline
\multicolumn{7}{c}{Galaxies thought not to be merging} \\
\hline
GEMS 033139.25m275303.0 & 03 31 39.25 & $-$27 53 03.0 & 0.20 & $-$21.5 & 1.1 & 00000 \\
GEMS 033139.03m275300.1 & 03 31 39.03 & $-$27 53 00.1 & 0.20 & $-$21.4 & 1.3 & 00000 \\
GEMS 033145.09m275520.6 & 03 31 45.09 & $-$27 55 20.6 & 0.49 & $-$20.9 & 0.9 & 03003 \\
GEMS 033145.09m275517.4 & 03 31 45.09 & $-$27 55 17.4 & 0.57 & $-$20.0 & 1.0 & 03003 \\
GEMS 033155.67m273840.0 & 03 31 55.67 & $-$27 38 40.0 & 0.519 & $-$20.8 & 1.2 & 03300 \\
GEMS 033155.83m273839.4 & 03 31 55.83 & $-$27 38 39.4 & 0.526 & $-$20.1 & 1.2 & 03300 \\
GEMS 033307.89m275150.4 & 03 33 07.89 & $-$27 51 50.4 & 0.518 & $-$21.6 & 1.3 & 00003 \\
GEMS 033307.75m275151.2 & 03 33 07.75 & $-$27 51 51.2 & 0.44 & $-$20.3 & 1.0 & 00003 \\
\hline
\multicolumn{7}{c}{Contentious cases} \\
\hline
GEMS 033117.77m275812.1 & 03 11 17.77 & $-$27 58 12.1 & 0.58 & $-$21.0 & 1.0 & 30303 \\
GEMS 033117.82m275814.1 & 03 11 17.82 & $-$27 58 14.1 & 0.55 & $-$20.3 & 1.0 & 30303 \\
GEMS 033313.59m275735.3  & 03 33 13.59 & $-$27 57 35.3 & 0.66 & $-$23 & 1.25 & 20310 \\
GEMS 033313.51m275737.3  & 03 33 13.51 & $-$27 57 37.3 & 0.65 & $-$22.5 & 1.1 & 20310 \\
GEMS 033123.29m274544.9A & 03 31 23.29 & $-$27 45 44.9 & 0.55 & $-$20.8 & 0.95 & 22101 \\
GEMS 033123.29m274544.9B & \nodata     & \nodata       & 0.55 & $-$19.3 & 0.95 & 22101 \\
\hline
\multicolumn{7}{c}{Mergers of galaxies with significant bulges and gas} \\
\hline
GEMS 033244.06m275143.3 & 03 32 44.06 & $-$27 51 43.3 &    0.26 &     $-$21.7 &        1.3 & 33333 \\
GEMS 033244.27m275141.1 & 03 32 44.27 & $-$27 51 41.1 &    0.273 &     $-$20.5 &       0.95 & 33333 \\
GEMS 033215.97m274943.1 & 03 32 15.97 & $-$27 49 43.1 &    0.59$^1$ &     $-$22.3 &        1.5 & 33033 \\
GEMS 033216.16m274941.6 & 03 32 16.16 & $-$27 49 41.6 &    0.67 &     $-$22.1 &       0.9 & 33033 \\
GEMS 033151.21m274659.3 & 03 31 51.21 & $-$27 46 59.3 &    0.67$^2$ &     $-$22.7 &       1.0 & 23333 \\
GEMS 033151.37m274700.3 & 03 31 51.37 & $-$27 47 00.3 &    0.64 &     $-$21.9 &        1.2 & 23333 \\
\hline
\multicolumn{7}{c}{Galaxies classified as dry mergers by the majority of classifiers} \\
\hline
GEMS 033136.86m280128.0A & 03 31 36.86 & $-$28 01 28.0 & 0.15 & $-$22.1 & 1.5 & 23222 \\
GEMS 033136.86m280128.0B & \nodata     &  \nodata      & 0.15 & $-$20.6 & 1.5 & 23222 \\
GEMS 033157.38m274239.0A & 03 31 57.38 & $-$27 42 39.0 & 0.62 & $-$22 & 1.3 & 22222 \\
GEMS 033157.38m274239.0B & \nodata     &  \nodata      & 0.62 & $-$22 & 1.3 & 22222 \\
GEMS 033208.85m275851.8$^3$ & 03 32 08.85 & $-$27 58 51.8  & 0.54 & $-$20.4 & 1.0 & 23321 \\
GEMS 033208.76m275851.2$^3$ & 03 32 08.76 & $-$27 58 51.2  & 0.54 & $-$20.5 & 1.0 & 23321 \\
GEMS 033210.77m274234.4A$^4$ & 03 32 10.77 & $-$27 42 34.4 & 0.419 & $-$21.9 & 1.2 & 22222 \\
GEMS 033210.77m274234.4B$^4$ & \nodata     &  \nodata      & 0.419 & $-$21.9 & 1.2 & 22222 \\
GEMS 033239.48m275301.6  & 03 32 39.48 & $-$27 53 01.6 & 0.65 & $-$22.5 & 1.25 & 22221 \\
GEMS 033239.47m275300.5  & 03 32 39.47 & $-$27 53 00.5 & 0.65 & $-$21.8 & 1.2 & 22221 \\
GEMS 033300.99m275545.2A$^5$ & 03 33 00.99 & $-$27 55 45.2 & 0.620 & $-$21.7 & 1.0 & 23321 \\
GEMS 033300.99m275545.2B & \nodata     &  \nodata      & 0.620 & $-$21.7 & 1.0 & 23321 \\
\hline
\vspace{-1.3cm}
\tablecomments{
$^1$ The redshift difference between GEMS 033215.97m274943.1 and 
GEMS 033216.16m274941.6 is somewhat larger than expected for COMBO-17's typical 
redshift error; either one of the redshifts is substantially in error
or our visual classification is inappropriate. \\
$^2$ The redshift difference between GEMS 033151.37m274700.3
and GEMS 033151.21m274659.3 is consistent with COMBO-17's redshift error 
(the galaxies' visual morphologies show that they are interacting, and 
are therefore at the same redshift). \\
$^3$ This merger candidate is one COMBO-17 object but was successfully 
split by the GEMS source extraction algorithm.  The coordinates are 
taken from GEMS, the redshift and color are adopted directly from COMBO-17, 
and the absolute magnitude of the COMBO-17 object is split into both galaxies
using the observed F850LP magnitude ratio from GEMS. \\ 
$^4$ GEMS 033210.77m274234.4A and B have been recently discussed in a
Keck laser guide star adaptive optics study of mergers detected 
by Chandra \citep[their XID-536; with an X-ray luminosity
of nearly $10^{42}$ ergs/sec]{melbourne05}.  Their luminosities and redshifts
are consistent with our own.  They fit model SEDs to the observed
0.4--2.2{\micron} broad band fluxes, concluding that both components
are dominated by old stellar populations at the epoch of observation. \\
$^5$ GEMS 033300.99m275545.2A and B have a COMBO-17 redshift of 0.45; 
the VVDS redshift is adopted in this paper.  Absolute magnitudes and 
rest-frame colors have been re-calculated using the correct VVDS
redshift. \\
GEMS galaxy names are simply a combination 
of their RA and Dec (where `m' denotes negative declinations), and 
A and B refer to the two sub-components when the system was not 
resolved into two separate galaxies by COMBO-17.  
Rest-frame $M_V$ values are approximate, split using 
our rough estimate of the luminosity ratio of the merger.  Redshifts
are accurate to $\delta z \sim 0.02$, magnitudes to $\sim 0.2$\,mag, 
and colors to $\sim 0.15$\,mag.  Redshifts quoted to 3 significant
figures are spectroscopic redshifts from the VVDS \citep{lefevre04}.
See the text for details of the classification scheme.  }
\end{tabular}
\end{center}
\end{table*}

For the sake of clarity, we present our selection criteria
for candidate dry mergers here.  Our philosophy is to 
catch the merging galaxies during the early stage of merging,
where the two galaxy nuclei are distinct entities allowing
relatively accurate characterization of the progenitors, 
including, importantly, an estimate of the 
mass ratio of the merger.  

Therefore, we select for further study very close pairs
of red-sequence bulge-dominated galaxies with  
{\it i)} projected separation $< 20$\,kpc, 
{\it ii)} a $V$-band absolute magnitude difference of 1.5\,mag or 
less (corresponding to a 4:1 luminosity ratio difference, our 
threshold for a major merger), {\it iii)} photometric redshift difference
$| \Delta z | < 0.1$.

However, not all projected close pairs will be merging;
some may be projections of galaxies at significantly different
redshifts, and some may be unbound fly-by interactions which will not merge.
In order to address these problems, it was necessary to add 
a second layer of selection.  Prolonged experimentation with 
automated merger diagnostics did not yield satisfactory results, and 
were unable to robustly flag the existence of broad, low surface brightness
sheets of debris which are the hallmark of a gas-poor major merger.
Therefore, we have used visual classification to screen 
the automatically-selected subsample of close pairs for signs
of dry merging.  This second level of visual selection is very stringent:
in demanding low-level tidal features or very close separations,
many merging systems are thrown out of the sample in phases where
the tidal features are unobservably weak (simulations suggest this 
should be the case around $\sim 1/2$ of the time).  
While this reduces the sample size, it provides some measure of security
against projections and fly-bys, and results in a smaller but cleaner
sample of dry merger candidates (spectroscopy of our dry merger candidates
is being sought and will be discussed in future works).

In order to calibrate out and understand
the subjectivity of visual classifications, we have
adopted a dual strategy: we have visually classified a suite of
simulated dry mergers and fly-bys(selected to have projected separation $\le 20$\,kpc)
in order to understand better
the signatures of a dry merger and to characterize the timescale
over which the signs of dry merging are recognizable (\S \ref{sims}), 
and we use 5 independent sets of classifications to quantify the 
degree of reproducibility of dry merger classifications (\S \ref{seln}).

\subsection{Expected Morphological Signatures of Dry Mergers} \label{sims}

In order to explore the expected appearance of morphological indicators of dry
mergers, and the timescales over which these indicators
are visible with single-orbit {\it HST} imaging, we make use of the simulations of 
\citet{naab05}.  The progenitors are early-type galaxies, and are themselves
the remnants of gas-free disk--disk 
mergers from \citet{naab03}\footnote{Gas-rich 
major merger remnants are somewhat diskier
than remnants from gas-free major disk mergers 
\citep[Naab \& Burkert, in prep.]{barnes02}.
The adoption of a gas-rich major merger remnant as the spheroid
merger progenitor in what
follows does not significantly affect the outcome, while
avoiding extra uncertainties associated with the detailed choice of 
star formation rate and feedback prescriptions. }.  
There are 400000 particles (160000 stars and 
240000 dark) per progenitor.  A number of simulations with 
different pericenters and orbital parameters (initially marginally bound, 
parabolic, or a hyperbolic fly-by) were carried out for two mass
ratios: 1:1 and 3:1.  A variety of different viewing angles 
were chosen between face-on and edge-on (with respect to the orbital plane).
The simulation output was recorded every 26\,Myr.
The simulations were used to construct artificial 
GEMS images; these were created by smoothing to GEMS resolution, 
re-binning to the appropriate angular diameter distance, and 
scaling to reproduce a given total luminosity of the 
remnant.  The images were then added to real sky background
from GEMS.

From examination of these simulated images (e.g., Fig.\
\ref{fig:sims}), some important points become clear.  The
morphological signatures of interactions between early-type galaxies
are often very weak; tidal tails are broad with low surface brightness
owing to the high velocity dispersion of the progenitor.  Furthermore,
there are nearly order-of-magnitude variations in overall merger
timescales (i.e., the time from first pass to final coalescence),
which depend sensitively on parameters such as mass ratio, pericenter
distance and orbit.  Importantly, however, the timescale over which
one can recognize tidal features in dry mergers is largely independent
of these parameters, and is comparable with the galaxies' internal
dynamical times.  Tidal features of sufficient strength to be detected
in our images occur only when the galaxies are very close together ---
typically between the last close pass prior to coalescence and
coalescence itself\footnote{In simulations with small pericenter
distances tidal features are visible for a short time after first
pass, but become rapidly too weak to observe.}.  Orientation plays a
role: interactions viewed nearly parallel to the orbital plane are
visible as very close pairs for 2--3 times longer than the same
interactions viewed from $< 60${\arcdeg} from face-on.

In order to estimate timescales over which GEMS-depth data
will be sensitive to dry mergers,
we explored the visibility of a wide range of early-type
galaxy interactions at a number of redshifts, and assuming
a range in final merger remnant luminosity.  
Preliminary explorations showed that luminous 
dry mergers at $z > 0.7$ are visible
only as a very close pair (timescales $\la 50$\,Myr, except 
for rare projections along the orbital plane); the
data lack sufficient depth to robustly detect tidal features.
Accordingly, we limit this study to $z<0.7$ galaxies and simulations.

Simulation snapshots were selected randomly from a variety
of interaction sequences; the only selection criterion applied 
was that the nuclei of the galaxies should be separated
by $<20$\,kpc (this is an analogous selection to that
applied to the data).  A total of 174 snapshots from 15 simulations with
different mass ratios (1:1 and 3:1), orbital parameters (marginally bound, 
parabolic, and a hyperbolic fly-by) and viewing angles (viewing 
angles from 0$\arcdeg$ from the normal to the orbital plane, 
30$\arcdeg$, 60$\arcdeg$, 80$\arcdeg$, and the edge-on 90$\arcdeg$ orientation)
met this separation criterion\footnote{Only one of the 
fly-by snapshots had separation $\le 20$\,kpc, reflecting the rapidity of 
fly-bys.}.  The random selection was weighted to ensure that 
the different viewing angles were selected approximately uniformly, 
in terms of solid angle.  Given that a simulation 
snapshot is created every
26\,Myr, this implies that the simulations have two distinct nuclei
separated by $<20$\,kpc for on average 300\,Myr.
The snapshots were scaled to a range of luminosities
$M_V < -20.5$ and redshifts $0.1<z<0.7$\footnote{The redshifts were 
drawn from a uniform deviate in terms of volume, not redshift, and 
are therefore weighted towards higher redshift in exactly the same 
way as the observational dataset (the average redshift of the simulations
was 0.53, whereas that of the observational dataset is 0.5).}, and a   
subsample of 25 snapshots were SExtracted\footnote{Some 
very close pairs were deemed to be one object by SExtractor
and were therefore fit with a single S\'ersic profile.}, GALFITted and
classified by the five classifiers (EFB, TN, DHM, CW and SK) in exactly the
same way as the data.
The classification was blind: simulation 
snapshots were intermixed with real dry merger candidates from GEMS.

Systems were deemed to be not merging if 2 classifiers or 
fewer classified the system
as a merger, possibly merging if 3 classifiers argued it was a merger, and 
merging if 4 or 5 classifiers decided it was a merging system.
According to this scheme, 12/25 snapshots were deemed not to be merging
and 13/25 were declared merging 
(there were no cases in which only 3 classifiers 
declared a system to be merging).  
Examples
of each class are shown in Figs.\ \ref{fig:sims_no} and 
\ref{fig:sims_mrg}.  It is apparent that snapshots with
very weak/nonexistent tidal features were typically
classified as non-interacting, while snapshots with either
{\it i)} $< 5$\,kpc separation and/or {\it ii)} 
tidal distortions (tails or asymmetries) were classified
as interacting.
Accordingly, we assign a timescale
of $\sim 150 \pm 50$\,Myr over which dry mergers are both automatically 
selected (with separations $< 20$\,kpc; timescales $\sim 300$\,Myr)
 and visually classified as
a merger (13/25 of the snapshots, plus associated counting uncertainties). 

\subsection{Selection of Dry Merger Candidates} \label{seln}

The dry merger simulations indicated that dry merger
identification is a well-posed problem with quantifiable
selection effects for galaxies with $z < 0.7$ and $M_V < -20.5$:
at larger redshift and/or fainter limits tidal features are no longer
easily visible in single-orbit depth HST/ACS imaging.
Accordingly, we adopt as the parent sample the sample
of red sequence galaxies with $U-V > 1-0.31z - 0.08(M_V + 20.77)$,
$0.1 \le z \le 0.7$, and $M_V < -20.5$ --- a sample of 468 galaxies.
Galaxy morphologies were 
assigned by eye for these galaxies: see e.g., 
\citet{mcintosh05}, \citet{wolf05}, or \citet{bellgems}
for a description of the morphological
typing criteria and for some examples from each type.
A total of 379/468 galaxies are classified to have 
early morphological types, E/S0/Sa\footnote{Systems 
with small amounts of ongoing star
formation in a weak disk component were allowed into the sample, 
as long as they were on the red sequence and their light distributions
were dominated by a smooth bulge component}.

Adopting the 468-galaxy red sequence $0.1 \le z \le 0.7$ and $M_V < -20.5$
sample as the parent sample, we have selected all close pairs satisfying
the following criteria: {\it i)} projected separation $< 20$\,kpc, 
{\it ii)} a $V$-band absolute magnitude difference of 1.5\,mag or 
less (corresponding to a 4:1 luminosity ratio difference, our 
threshold for a major merger), {\it iii)} photometric redshift difference
$| \Delta z | < 0.1$.  Ten pairs were found with these properties.

However, from the visual classifications there were a number 
of merging systems which were classified by \combo as one object.
It is important to augment the sample of merger candidates with
such blended mergers.  Every one of the 468-galaxy parent sample
has been fit using the GALFIT galaxy fitting code \citet{peng02}
using a single S\'{e}rsic profile\footnote{The
\citet{sersic68} model has a profile with
surface brightness $\Sigma \propto e^{-r^{1/n}}$, where $r$ is the
radius and $n$ is an index denoting how
concentrated the light distribution of a given galaxy is:
$n = 1$ corresponds to an exponential light distribution, and
$n = 4$ corresponds to the well-known de Vaucouleurs profile.}.  
A sample of 34 galaxies was
automatically selected which had strong residuals from the 
S\'ersic profile fit (within one half-light radius, 
the sum of the absolute values of the residuals exceeded
$15$\% of the total flux within the half-light radius).
The vast majority of these 34 systems were 
bulge/disk systems or had strong dust features and lanes:
only 6 systems had evidence for multiple nuclei. 
These 6 systems were added to the sample of 10 COMBO-17-selected
close galaxy pairs, forming the final sample of 16 dry merger
candidates.  Postage stamps and GALFIT residuals
of all 16 candidates can be found in 
Figs.\ \ref{fig:nointeract}--\ref{fig:drygals}.

\vspace{1cm}

\section{Results} \label{res}

\begin{figure}[tbh]
\epsfbox{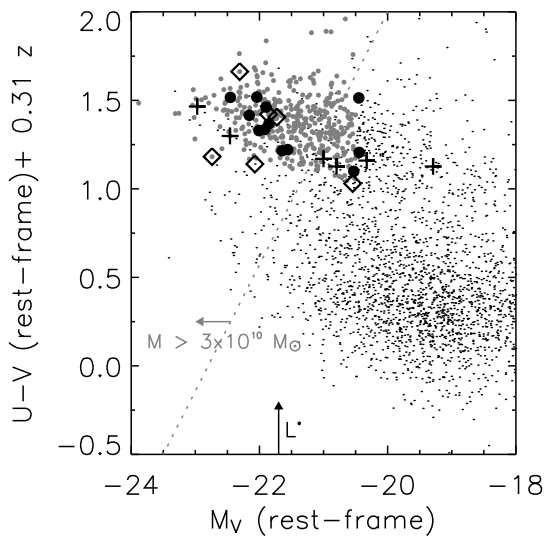}
\caption{\label{fig:cmr} The distribution of the candidate 
spheroid mergers in the color--magnitude diagram.  Points denote
all galaxies with $0.1<z<0.7$; grey circles early-type galaxies
on the red sequence.  Black symbols show 
possible mergers (crosses), gas-rich mergers (diamonds), and 
dry mergers (filled circles).
In the cases where \combo 
was unable to resolve the merger from the ground and assigned
a single redshift, we split the flux between the 
two galaxies following Table \ref{tab} and 
assign the same color; small random offsets
are added to aid visibility.  The arrow shows the approximate
position of $L^*$ for red sequence galaxies at $z \sim 0.5$, 
and the dotted line denotes the locus of galaxies
of constant stellar mass $\sim 3\times 10^{10} M_{\sun}$, 
assuming a \citet{kroupa01} stellar initial mass function.
}
\end{figure}

The 16 dry merger candidates were classified
by the 5 classifiers (EFB, TN, DHM, CW and SK), with 
4 different classification categories: (0) no evidence
for an interaction, (1) possible merger, (2) dry merger, and 
(3) gas-rich merger.  
The results from all classifiers
are given in Table \ref{tab}.  On the whole, agreement between 
the classifiers was good, and the galaxies separated reasonably
cleanly into 4 subsamples: galaxies where the majority of classifiers
felt that there was no evidence for merging (Fig.\ \ref{fig:nointeract}),
systems where classification was uncertain (Fig.\ \ref{fig:poss}),
gas-rich mergers (Fig.\ \ref{fig:gas_rich}), and 
dry mergers (Fig.\ \ref{fig:drygals}).  A firm classification 
as a merging system required that 4 or 5 classifiers felt that 
a given system was a definite merger.  A total of 6 systems were 
identified as dry mergers, three were identified as gas-rich mergers
(these systems all had prominent bulge components and red colors), 
and 2 of the possible mergers could qualify as a dry merger
as both components were bulge-dominated (GEMS 033123.29m274544.9
and GEMS 033313.59m275735.3).  Fig.\ \ref{fig:cmr} 
shows the color-magnitude diagram of all
 $0.1<z<0.7$ galaxies from \combo 
(points), all red-sequence early-type galaxies (gray circles), 
and the merger candidates (black symbols, where symbol type
denotes classification).  

On this basis, it is possible to estimate the incidence of major
mergers between luminous early-type galaxies from $z \sim 0.7$ to the
present day.  Among the 379 red, early-type galaxies in GEMS with $M_V
< -20.5$ and $0.1<z<0.7$, we find a total of 12 galaxies (6 pairs)
that are strong dry major merger candidates.  
Classifications of dry merger simulation snapshots suggest
that we will be
able to recognize a dry merger for roughly an internal dynamical time,
or approximately 150\,Myr.  Assuming a constant dry merger fraction
over the 6.3\,Gyr interval $0<z<0.7$, an average, luminous $M_V <
-20.5$ early-type galaxy will undergo $(6.3/0.15) \times (12/379) \sim
1.3_{-0.4}^{+0.7}$ dry major merger (with mass ratios between 1:1 and
4:1).  Yet, because there are only 6 merging pairs, we cannot strongly
constrain the merger fraction evolution. 
The evolution of the merger fraction has been argued 
to evolve as rapidly as $(1+z)^3$ \citep[e.g.,][]{lefevre00}.
If the dry merger fraction were to evolve similarly, 
an average luminous $M_V < -20.5$
early-type galaxy will undergo $\sim 0.9_{-0.3}^{+0.5}$ major dry merger between 
$z = 0.7$ and the present day.  Obviously, our limited number
statistics are currently a dominant source of uncertainty;
nonetheless, our results indicate that dry merging plays an important
role in shaping the properties of luminous early-type galaxies and
building up the mass contributed by red-sequence galaxies since $z\sim1$.

\section{Discussion}  \label{disc}

\begin{figure}[tbh]
\epsfbox{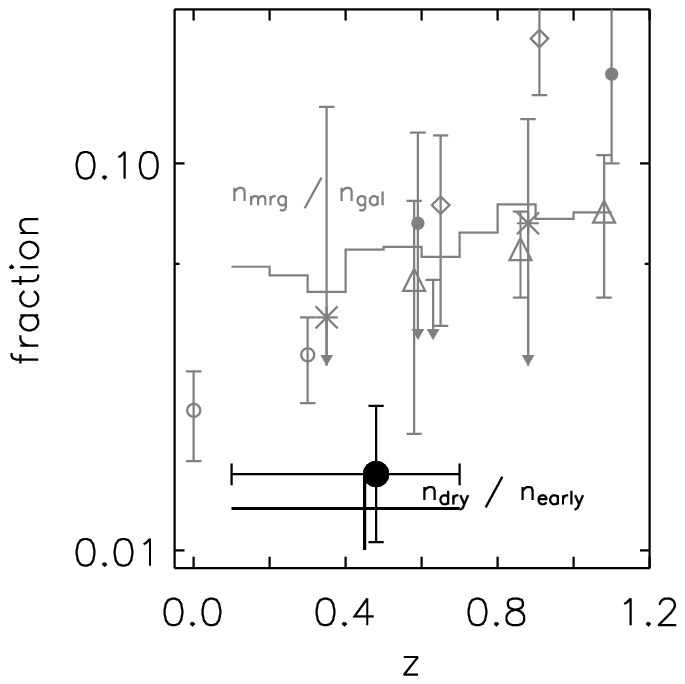}
\caption{\label{fig:model} A comparison of the dry merger
fraction with previous observations of merger fractions, 
and the predictions of a semi-analytic model.
Gray data points show the merger fraction of galaxies with 
$M_B \la -20$ from \citet[open circles]{patton02}, 
\citet[open diamonds]{lefevre00}, \citet[solid circles]{conselice03},
\citet[triangles]{lin05}, 
and \citet[crosses, and the upper limit at $z=0.63$]{bundy03}.
The model prediction of the major merger rate for $M_B < -20$ 
galaxies is shown 
as a solid gray line for the redshift range $0.1<z<1.1$.  
The black data point shows the 
dry merger fraction from this work.  The black naked error
bar shows the model prediction for the fraction of 
$M_V < -20.5$ red early-type
galaxies that have experienced a major merger in the 
last 150\,Myr averaged over $0.1<z<0.7$
with associated number uncertainties.   }
\end{figure}

\subsection{Are our observations consistent with other 
observational constraints?}

Is our suggestion that nearly every luminous early-type galaxy in the
nearby universe has experienced a major gas-free merger since $z\sim1$
consistent with existing observational constraints?  Dissipationless
($=$gas-free) merging preserves the fundamental plane
\citep[e.g.,][]{nipoti03,gonzalez03,boylan05}; therefore the thickness of the
fundamental plane is not a strong constraint on the prevalence of dry
merging.  In contrast, simulations and models have demonstrated that
dry merging will lead to scatter in the stellar mass--size relation
\citep[equivalently, the Kormendy relation;][]{nipoti03} and the
color--magnitude relation \citep[CMR;][]{bower98}.  Observations of
the evolution and scatter of these scaling relations limit the amount
of dry merging to around a factor of two increase in mass since $z
\sim 1$ \citep{mcintosh05,bower98}\footnote{The CMR limit applies
since the epoch when the CMR was imprinted on the early-type galaxy
population.}.  A similar limit is derived from the stellar mass
evacuated from the cores of luminous elliptical galaxies by their
merging black holes in the absence of gas \citep{graham04}.

\subsection{Comparison with previous measurements of merger fraction}

It is interesting to compare our results with previous observational
estimates of the overall galaxy merger fraction.  In Fig.\
\ref{fig:model}, we show the dry merger fraction (black solid point)
of $M_V < -20.5$ red early-type galaxies, along with the fraction of
$M_B \la -20$ galaxies in close pairs with $\le 20$\,kpc separation
taken from \citet[open diamonds]{lefevre00} and
\citet[open circles]{patton02}.  The results for \citet{lin05}
are shown as triangles (separations between 10 and 30\,kpc; 
$M_B \la -20$).
\citet{bundy03} argued that 
many close pairs with nearly equal optical magnitudes were minor mergers
where the satellite galaxies had boosted 
optical flux from tidally-induced bursts of star formation;
they used near-infrared photometry to estimate the major merger
fraction (crosses, and the upper limit at $z=0.63$).
The solid circles denote the merger fraction of $M_B < -20$ galaxies,
inferred using morphological disturbance as an indicator of past
merging activity \citep{conselice03}. 

It is clear that the dry merger fraction is lower than 
the general merger fraction.  A number of factors may contribute
to this reduction. Close pairs and morphologically-selected
gas-rich mergers are expected to be visible for
a timescale $\tau \sim 0.5-1$\,Gyr \citep{lefevre00,patton02,conselice03},
whereas dry mergers are visible for $t_{\rm dry} \sim 150$\,Myr.  
Episodes of star formation in gas-rich mergers
will increase the brightness of the merging galaxies, increasing
their prominence relative to the star formation-free dry 
mergers.  Finally, gas-rich galaxies tend to inhabit 
lower-density environments than gas-poor galaxies;
this difference in environment will also
affect the likelihood of a major merger.  

\subsection{Theoretical Expectations in a Hierarchical Universe}

Do hierarchical models predict that dry mergers are an important mode
of transformation and mass growth for luminous elliptical galaxies?
To address this question, we make use of an updated version of the
Somerville et al.\ semi-analytic galaxy formation model
\citep[see][for a description of the basic ingredients]{sp,spf}, based
in the standard $\Lambda$CDM paradigm of hierarchical structure
formation. The models are based on Monte Carlo realizations of dark
matter halo merger histories, constructed using the method described
in \citet{sk:99}, and include fairly standard recipes
treating gas cooling, star formation, and supernova feedback. The star
formation rate is given by $\dot{m}_{*} = m_{\rm cold}/(\tau^0_{*}
t_{\rm dyn})$, where $m_{\rm cold}$ is the cold gas mass in the
galaxy, $t_{\rm dyn}$ is the dynamical time of the disk, and
$\tau^0_{*}$ is a free parameter, which is adjusted to obtain
agreement with observed gas fractions in spiral galaxies at the
present day. We convolve the resulting star formation histories with
the multi-metallicity stellar SED models of \citet{bc:03}, and include
the effects of dust extinction using a simple relationship between
face-on B-band optical depth and star formation rate ($\tau_B \propto
SFR^\beta$), based on the observational results of
\citet[we assume $\beta=0.5$]{wangheckman:96}. Merging of galaxies
(sub-structure) within virialized dark matter halos is tracked by
computing the time it takes for a satellite galaxy to lose all of its
orbital angular momentum via dynamical friction against the background
of the dark matter halo, using the standard Chandrasekhar
approximation. Mergers with mass ratio 1:4 or greater are assumed to
result in a burst of star formation, and all pre-existing 'disk' stars
are transferred to a 'bulge' component.

Earlier versions of this model, like many other models in the
literature, had difficulty producing enough luminous red galaxies,
particularly at high redshift \citep[see, e.g.,][]{somer04}, and
therefore could not produce a useful prediction of the merger rate for
these systems.  Somerville et al. (in prep.) find that introducing an
ad hoc `quenching' of star formation when the mass of the bulge
component grows larger than $M_{\rm bulge,crit} \simeq 2 \times
10^{10} M_{\sun}$ leads to much improved agreement with the observed
$0 < z \la 1 $ color distributions as a function of absolute
magnitude, including the observed bimodality of the color
distributions at $z\sim0$--2.  A full description of this model will
be presented in this forthcoming work: at this stage, we argue that as
the model approximately reproduces the global luminosity functions and
observed number of red, early type galaxies at $z\sim0$--1, it is at
least a useful starting point to examine the overall merger rate and
the dry merger rate.

In Fig.\ \ref{fig:model}, we show the predicted fraction of $M_B \la
-20$ galaxies that have experienced a major merger within the past
1\,Gyr (the grey line; major mergers are defined as having mass
ratios between 1:1 and 4:1).  
These predictions are derived from a mock light cone from $0.1<z<1.1$ 
covering a sky area of 1 sq.\ deg.\ 
(corresponding to about four times the volume of GEMS).
Bearing in mind on one hand the difficulties of
measuring galaxy merger rate from close pairs, and on the other hand
uncertainties in physical prescriptions for dynamical friction, tidal
stripping of galaxy halos, and tidally-induced star formation,
the agreement between the model and observations
is rather good.   There is a hint that the model merger fraction 
decreases rather less rapidly than the observations;
interestingly, this behavior is also seen in the semi-analytic
models of \citet{benson02}, as shown in Fig.\ 13 of 
\citet{conselice03}.  

The black naked error bar shows the dry merger fraction predicted by
this model.  We identify dry mergers in the model as galaxies with
magnitudes $M_V < -20.5$, $U-V$ colors on the red sequence, and which
have had a major merger in the past 150\,Myr.
The mock catalog has roughly four times the volume of GEMS; there
are 13 merging systems from 982 early-type 
$M_V < -20.5$ red sequence galaxies in this mock catalog
in the interval $0.1<z<0.7$.  The inferred merger fraction 
is in excellent quantitative agreement with 
our observational determination (6 systems from 379 galaxies).
This result lends further theoretical support
to hydrodynamical studies of elliptical galaxy formation
\citep{dominguez04} and semi-analytic
studies of galaxy clusters \citep{khochfar03}, both of which 
argue that dry mergers are an important process in
driving the buildup of massive early-type galaxies at recent times.

\section{Conclusions} \label{sec:con}

We have used the GEMS and COMBO-17 surveys in conjunction with
$N$-body and semi-analytic galaxy formation simulations to explore the
frequency of gas-free major mergers between spheroid-dominated
galaxies (dry mergers) since $z = 0.7$.  We focused on this aspect of
galaxy merging both because identifying gas-free mergers is less
likely to be complicated by the effects of merger-induced star
formation and the accompanying dust extinction,
and because it can be modeled relatively 
straightforwardly (for similar reasons).
The morphological signatures of such interactions were
calibrated using mock GEMS-like images based on $N$-body simulations, and include 
$< 5$\,kpc separation of close nuclei and/or broad tidal tails and asymmetries.
The simulations showed that such features are visible only for
galaxies with $M_V \la -20.5$ in single-orbit {\it HST} F850LP data 
at $z \la 0.7$; higher-redshift
galaxies show unobservably weak tidal features in the GEMS data.
A total of 809 red-sequence galaxies were visually-inspected 
for morphological signatures of spheroid merging; 6 systems (12 galaxies) 
with luminosity ratios between 1:1 and 4:1 were found
out of a total of 379 red sequence early-type galaxies with $M_V < -20.5$.
The simulations suggest that morphological signatures of spheroid
merging are visible for $\sim 150$\,Myr; we therefore argue that an
average $M_V < -20.5$ early-type galaxy has experienced between 0.5
and 2 major spheroid mergers since $z \sim 0.7$.  This merger frequency
is consistent with limits of $\la 1$ major gas-free merger at recent
times, as derived from the color--magnitude and stellar mass--size
relations, and from core mass deficits. We have compared this result
with an updated semi-analytic model.  This model reproduces the
evolution of the number of luminous red early-type galaxies and the
overall merger fraction reasonably well, and can therefore be used to
obtain a plausible prediction of the frequency of dry mergers in a
hierarchical universe.  The predicted dry merger fraction is
consistent with the observations to within their combined
uncertainties.  Thus, both observations and theory lend strong support
to the notion that major spheroid mergers are an important driver of
the evolution of luminous early-type galaxies in recent epochs.

\acknowledgements
We wish to thank the referee for a number of helpful suggestions.
E.\ F.\ B.\ wishes to thank Ralf Bender and Marijn Franx 
for interesting discussions. 
Based on observations taken with the NASA/ESA {\it Hubble Space
Telescope}, which is operated by the Association of Universities
for Research in Astronomy, Inc.\ (AURA) under NASA contract NAS5-26555.
Support for the GEMS project was provided by NASA through grant 
number GO-9500 from the Space Telescope Science Institute.
E.\ F.\ B.\ was supported by the European Community's Human
Potential Program under contract HPRN-CT-2002-00316 (SISCO).
S.\ J.\ and D.\ H.\ M.\ acknowledge support from NASA under
LTSA Grant NAG5-13063 and NAG5-13102 respectively.
C.\ W.\ was supported by a PPARC Advanced Fellowship.
K.\ J.\ was supported by the German DLR under project number
50~OR~0404.  C.\ H.\ acknowledges financial support
from the GIF.

\appendix

\section{A. Are fly-bys important?} \label{app}

\citet{tran05} recently showed that all luminous dry merger
candidates in the outskirts
of the $z \sim 0.8$ MS1054 galaxy cluster 
with sufficient S/N to measure velocity
offsets were indeed bound $R < 10$\,kpc pairs (with velocity
differences less than $200\,{\rm km\,s}^{-1}$).  Noting that 
\citet{tran05} and \citet{vandokkum99} argue that most 
of these galaxies are in the cluster infall region (and therefore
are in group environments which are about to be incorporated
into the cluster proper), their study supports the notion 
that the majority of the dry merger candidates studied
in this paper are bound and will likely merge.

Yet, one of the few published examples of an E--E interaction in the
local universe (NGC 4782/4783) has been shown by \citet{mad91} to have
an unbound hyperbolic orbit.  Based on simulated images from
simulations of unbound dry mergers similar to those discussed in the
main text, we find that morphological disturbances generated by fly-by
interactions are short-lived, $\ll 50$\,Myr for systems of a similar
luminosity to the ones we observe here. Nonetheless, if the number of
fly-by interactions is much larger than the number of mergers, fly-bys
could significantly contaminate a spheroid major merger sample.

Because galaxies with velocity differences greatly 
in excess of their internal velocity dispersion are 
extremely unlikely to merge, it is worth understanding 
the likely contribution of high-velocity fly-bys to 
tidally-disturbed spheroid-dominated systems in 
a cosmic-averaged sense.  Relative velocity
difference is a strong function of the velocity
dispersion of the local environment. We explore
the likelihood of producing 
a given tidal effect in different environments by adopting
the impulse approximation.  We wish to work out the number
of interactions as a function of environment, characterized
by the velocity dispersion of the local environment $\sigma$.
The number of interactions $N_{int} \propto n_{\rm E/S0}^2 r^2$, 
where $n_{\rm E/S0}$ is the number density of E/S0 galaxies and $r$
is the pericentric distance.  The dependence of cross-section 
on velocity can be estimated by 
assuming that the relative
velocity $v \propto \sigma$.  In order to 
work out the cross-section for interaction producing a given
displacement in stars $s$ as a function of 
$v$, we note that
tidal forces $F \propto r^{-3}$. 
Since interaction timescale $t \propto 1/v$, the displacement
in stars $s \propto F t^2 \propto r^{-3} v^{-2}$.  Thus, 
given a constant tidal displacement of stars 
$s$, the impact parameter $r \propto v^{-2/3}$.
The number of E/S0 galaxies can be estimated 
assuming a constant group stellar M/L; $n_{\rm E/S0} \propto f_{\rm E/S0} 
M_{\rm halo} / r_{\rm halo}^3$.  Since $M_{\rm halo} \propto \sigma^3$ and 
$r_{\rm halo} \propto \sigma$, $n_{\rm E/S0} \propto f_{\rm E/S0}$ alone; 
we assume for simplicity $f_{\rm E/S0} \propto \sigma \propto v$.  Thus,
$N_{int} \propto v^{2/3}$.  To estimate the fraction of galaxies with this
given displacement of stars owing to tidal affects, $f_{\rm int}$, 
one divides by the total number of E/S0 galaxies
$N_{E/S0} \propto f_{\rm E/S0} M_{\rm halo} \propto v^4$; thus, 
$f_{\rm int} \propto v^{-10/3}$.  A cosmic-averaged fraction 
must be further weighted by the halo mass function of group and cluster-sized
halos, $\Phi(\sigma) \propto \sigma^{-3}$.  Thus, despite the 
dominance of E/S0 galaxies in clusters, the vast majority
of tidally-disturbed spheroid-dominated galaxies are in the lowest
velocity-dispersion groups capable of hosting E/S0 galaxies, and 
are therefore likely to merge.  

A complementary analysis following \citet{makino97}, asking 
what the average environment of spheroid mergers should be (using the scalings
above), comes to a similar conclusion: $f_{\rm mrg} \propto 1/\sigma^2$, and 
weighting by the mass function of groups and clusters, the cosmic-averaged
fraction is again completely dominated by low-speed E/S0 mergers in 
low-mass groups.

\end{document}